\documentclass{iau} 
\usepackage{graphicx}

\title[Environment: star formation and AGN] 
{The role of environment on quenching, \\ star formation and AGN activity}

\author[Bianca M. Poggianti \& the GASP team]   
{Bianca M. Poggianti$^1$, Callum Bellhouse$^1$, Tirna Deb$^2$, Andrea
  Franchetto$^{1,3}$, Jacopo Fritz$^4$, Koshy George$^5$, Marco
  Gullieuszik$^1$, Yara Jaff\'e$^6$, Alessia Moretti$^1$, Ancla
  Mueller$^7$, Mario Radovich$^1$, Mpati Ramatsoku$^8$, 
Benedetta Vulcani$^1$
 \and the rest of the GASP team\thanks{http://web.oapd.inaf.it/gasp/index.html}}

\affiliation{$^1$INAF-Osservatorio Astronomico di Padova, vicolo
  dell'Osservatorio 5, 35122, Padova, Italy, 
\\ email: {\tt bianca.poggianti@inaf.it}
\\[\affilskip]
$^2$Kapteyn Astronomical Institute, University of Groningen, Postbus
800, NL-97009 AV, Groningen, The Netherlands, \\  
$^3$ Dipartimento di Fisica e Astronomia, Universit\'a di Padova,
vicolo dell'Osservatorio 3, 35122 Padova, Italy, \\
$^4$Instituto de radioastronomia y Astrofisica, UNAM, Campus Morelia,
A.P. 3-72, 58089, Mexico    , \\
$^5$Faculty of Physics, Ludwig-Maximilians-Universitat,
Scheinerstr. 1, 81679, Munich, Germany, \\
$^6$Instituto de Fisica y Astronomia, Universidad de Valparaiso, Avda.
Gran Bretana 1111, Valparaiso, Chile, \\
$^7$Ruhr University Bochum, Faculty of Physics and Astronomy,
Universitatsstr. 150, 44801 Bochum, Germany, \\
$^8$Department of Physics and Electronics, Rhodes University, PO Box
94, Makhanda, 6140, South Africa\thanks{INAF-Osservatorio Astronomico di Cagliari, via della Scienza 5,
09047 Selargius (CA), Italy}}

\pubyear{2020}
\volume{359}  
\jname{Galaxy evolution and feedback across different environments}
\editors{T. Storchi-Bergmann, R. Overzier, W. Forman \& R. Riffel, eds.}
\begin{document}

\maketitle

\begin{abstract}
Galaxies undergoing ram pressure stripping in clusters are an
excellent opportunity to study the effects of environment on
both the AGN and the star formation activity.
We report here on the most recent results from the GASP survey. We
discuss the AGN-ram
pressure stripping connection and some evidence for AGN feedback
in stripped galaxies. We then focus on the star formation activity, both in the
disks and the tails of these galaxies, and conclude drawing a picture
of the relation between multi-phase gas and star formation.
\end{abstract}

\firstsection 
\section{Introduction}

Spiral galaxies in clusters and groups lose their gas due to the ram
pressure exerted by the hot intergalactic medium on the galaxy
interstellar and circumgalactic medium. The effects of ram pressure
stripping (RPS) on the disk gas have been 
observed at several different wavelengths (e.g. Gavazzi 1989, Kenney et al. 2004, Yagi et
al. 2010, Sun et al. 2010, Smith et al. 2010, Ebeling et al. 2014,
Gavazzi et al. 2018, Boselli et al. 2020)  and have been predicted by both analytical
approaches and hydrodynamical simulations (Gunn \& Gott 1972, Tonnesen
\& Brian 2009, Roediger et al. 2008, 2014).  Stripped galaxies offer
a great opportunity to study several fundamental physical processes in
astrophysics, especially thanks to recent integral-field spectroscopic
studies.

Hereafter, we will discuss the latest results of ram pressure
studies concerning three fields of research: the triggering of AGN activity;
the star formation process within and outside of galaxy disks; and 
the baryonic cycle between multi-phase gas and star formation.
As we discuss below, unexpected findings were uncovered for each of
these fields.

\begin{figure}[b]
\begin{center}
\includegraphics[width=5.3in]{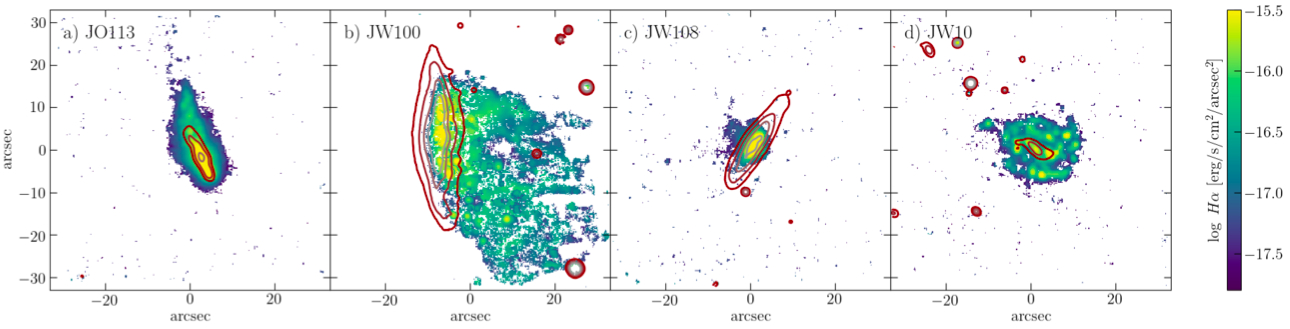}
\caption{MUSE $\rm H\alpha$ surface brightness maps for four GASP
  galaxies in different conditions of RPS: a galaxy undergoing moderate
  stripping (JO113); a jellyfish galaxy (JW100); an
  advanced stage of stripping, with gas left only in the central
  region of the disk (JW108) and a galaxy that is disturbed but not
  stripped (JW10): the latter is a merger, as testified by the stellar velocity map (not shown). 
Red contours delimit the galaxy stellar disk. From Jaff\'e et al. 2018.}
\label{fig1}
\end{center}
\end{figure}

Our summary is mostly based on results from the survey GASP (GAs Stripping
Phenomena in galaxies, Poggianti et al. 2017a,
http://web.oapd.inaf.it/gasp/index.html), 
which includes a MUSE integral-field ESO Large
Program and follow-up multiwavelength programs investigating the
molecular gas (APEX, ALMA), the neutral gas (JVLA, MeerKAT) and the
young stellar content (UVIT@ASTROSAT). The GASP sample includes
cluster galaxies at different stages and different strengths
of the stripping process (Fig.~1), from initial to peak stripping to
the late phases with little gas left, and even fully stripped
post-starburst galaxies and an undisturbed control sample. 
GASP also includes a group and field
subsample of galaxies, which are not discussed here (Vulcani et
al. 2017, 2018a,c, 2019a,b).

\section{AGN}
An unexpected result was the high incidence of AGN among the so
called ``jellyfish galaxies'', defined as galaxies with one-sided
tails of ionized gas (longer than the stellar disk diameter). 
MUSE data demonstrate that the tails are due to RPS. Six out of the seven GASP
jellyfish galaxies studied hosted an AGN (one of them is an optical
LINER). This AGN incidence is much higher than in general cluster and
field samples, suggesting that ram pressure can cause gas to flow towards the center
and trigger the AGN activity (Poggianti et al. 2017b, Fig.~2).

\begin{figure}[b]
\begin{center}
\centerline{\includegraphics[width=3.9in]{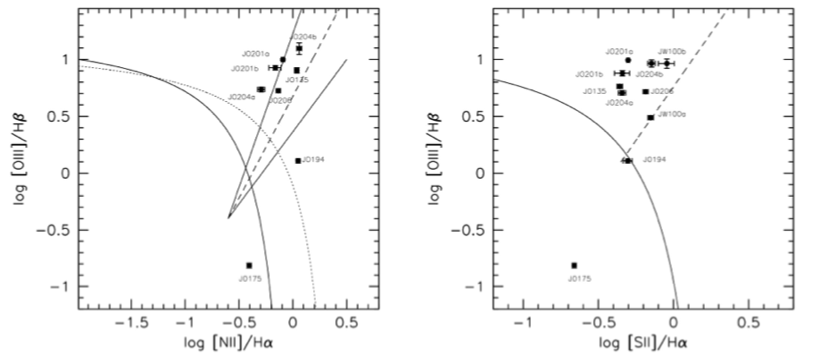}\hfill\includegraphics[width=1.6in]{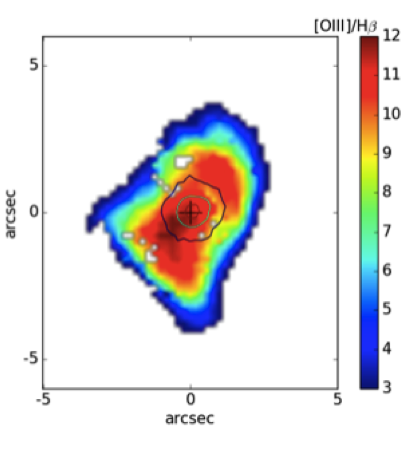}} 
\caption{Left and center. BPT line-ratio diagnostic diagrams for the jellyfish sample
  from Poggianti et al. (2017). Most galaxies lie in the AGN region of
  this diagram. Right: The high [OIII]/$\rm H\beta$ ratio of the central region
of the JO201 jellyfish galaxy indicates the presence of the AGN. The
black contour shows the region with emission of the coronal [Fe
VII]$\lambda$6087 line, also indicative of an AGN. From Radovich et
al. (2019).}
\label{fig1}
\end{center}
\end{figure}

The exact physical mechanism responsible for the gas inflow still
needs to be pinpointed. It may be due to a loss of
angular momentum of the galactic gas when it interacts with the
non-rotating intracluster-medium (Tonnesen \& Bryan 2012), or it can be
generated by oblique shocks in a disk flared by the magnetic field
(Ramos-Martinez et al. 2019).
Very recent high resolution simulations of a galaxy cluster also find that
ram pressure triggers enhanced accretion onto the central black hole (Ricarte et
al. 2020). 

In this context, it is relevant to ask: 
a) how sure is the presence of the AGN, and could the gas
ionization be due to shocks or other mechanisms? Based on the
comparison with AGN, shocks and HII-region photoionization models and
using different line ratios, Radovich et al. (2019) confirmed the univocal 
interpretation of the presence of
AGN. The same work found iron coronal lines (Fig.~2) and extended ($> 10 \rm kpc)$
AGN-powered ionization cones in some of these galaxies, as well as AGN 
outflows extending out to
1.5-2.5 kpc from the center, with outflow velocities in the range
250-550$\rm km \, s^{-1}$.
b) The sample published in Poggianti et al. (2017b) is small, and
consists of quite massive galaxies ($\geq 4 \times 10^{10}
M_{\odot}$). How significant is the enhancement of the AGN fraction,
and is that confirmed by further studies?
Is the RPS-enhanced AGN activity present only under certain
circumstances, e.g. in a certain stage of stripping (when it is
strongest), or for certain orbits within the cluster, etc? Or does it
occur only in galaxy clusters with certain intracluster medium properties?
For example, Roman-Oliveira et al. (2019), in their study of the A901/2
supercluster, find only 5 AGN host galaxies in their sample of 58
jellyfish candidates with an assigned classification (see also
Roman-Oliveira in these proceedings).
The analysis of the whole GASP sample is underway, and studies of
other samples/redshifts will help clarify this point, keeping in mind that 
the detection of a Seyfert2/LINER AGN depends crucially on the data
quality and sensitivity. Furthermore, for several galaxies there is
evidence for a large amount of dust in their nuclear regions: in this case,  
the optical line diagnostic ratios provided even by deep MUSE data
sometimes may not reveal the dust obscured AGN (e.g. Fritz et
al. 2017), and X-ray data would be required for its detection. 

\begin{figure}[b]
\begin{center}
\centerline{\includegraphics[width=1.5in]{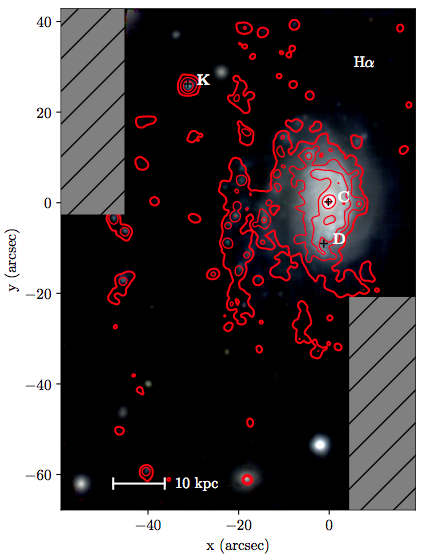}\includegraphics[width=3.6in]{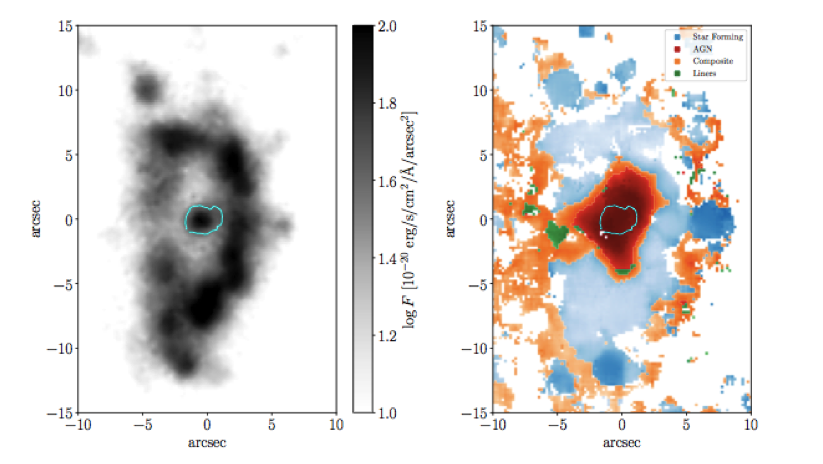}\hspace{-0.8cm}\includegraphics[width=2.1in]{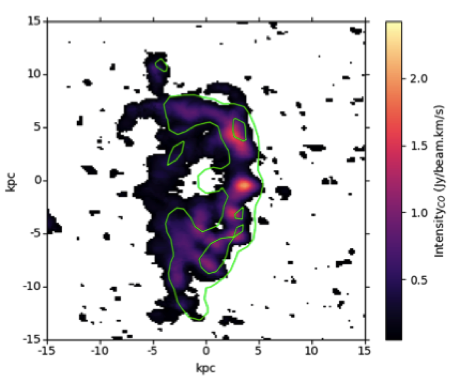}} 
\caption{Left. The jellyfish galaxy JO201 $\rm H\alpha$ contours
  superimposed on the stellar image. The long extraplanar tails of
  ionized gas are visible (Bellhouse et al. 2017, 2019). Other three
  panels: a zoom on the disk of (from left to right) NUV emission,
  ionization source map and CO map (George et al. 2019). The 8 kpc
  central hole in UV and CO emission corresponds to the AGN-powered
  $\rm H\alpha$ emission.}
\label{fig2}
\end{center}
\end{figure}

Moreover, the combination of MUSE and multiwavelength data has provided strong
evidence for the effects of AGN feedback in the jellyfish galaxy  
 JO201 (George et al. 2019, see also Bellhouse et al. 2017, 2019). The
 central 8kpc region of JO201 is depleted
 of both molecular gas (as traced by a CO ALMA observation) 
and of recent and ongoing star formation (as traced by NUV and FUV
imaging with UVIT@ASTROSAT) (Fig.~3). This region is filled with gas
ionized by the AGN (as seen by MUSE). Evidence for a similar effect in
other GASP jellyfish galaxies is present and is currently under investigation.

\section{Star formation}
The effects of RPS on the star formation activity are variegated and
in a sense counterintuitive, since RPS removes gas which is the
fuel for the formation of new stars.

\begin{figure}[b]
\begin{center}
\centerline{\includegraphics[width=2.3in]{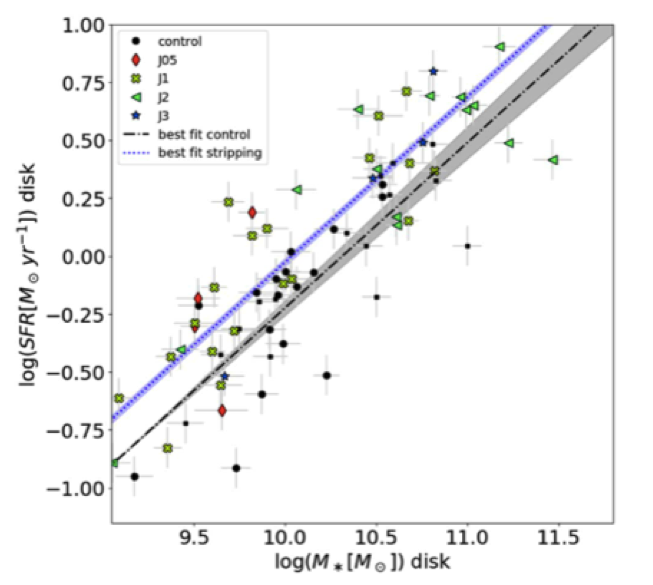}\hfill\includegraphics[width=3.0in]{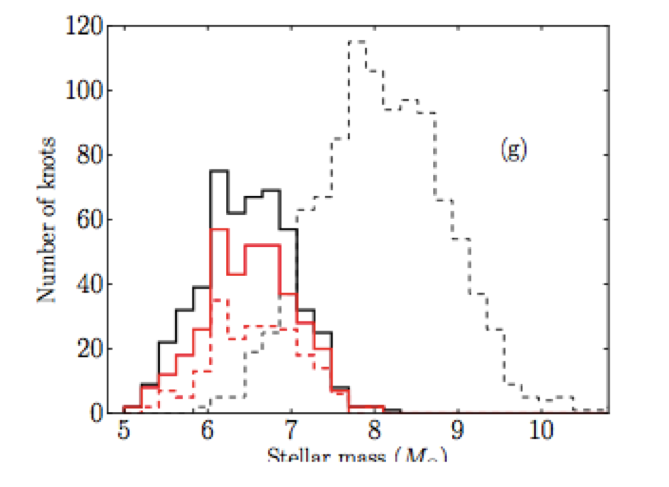}} 
\centerline{\includegraphics[width=4.5in]{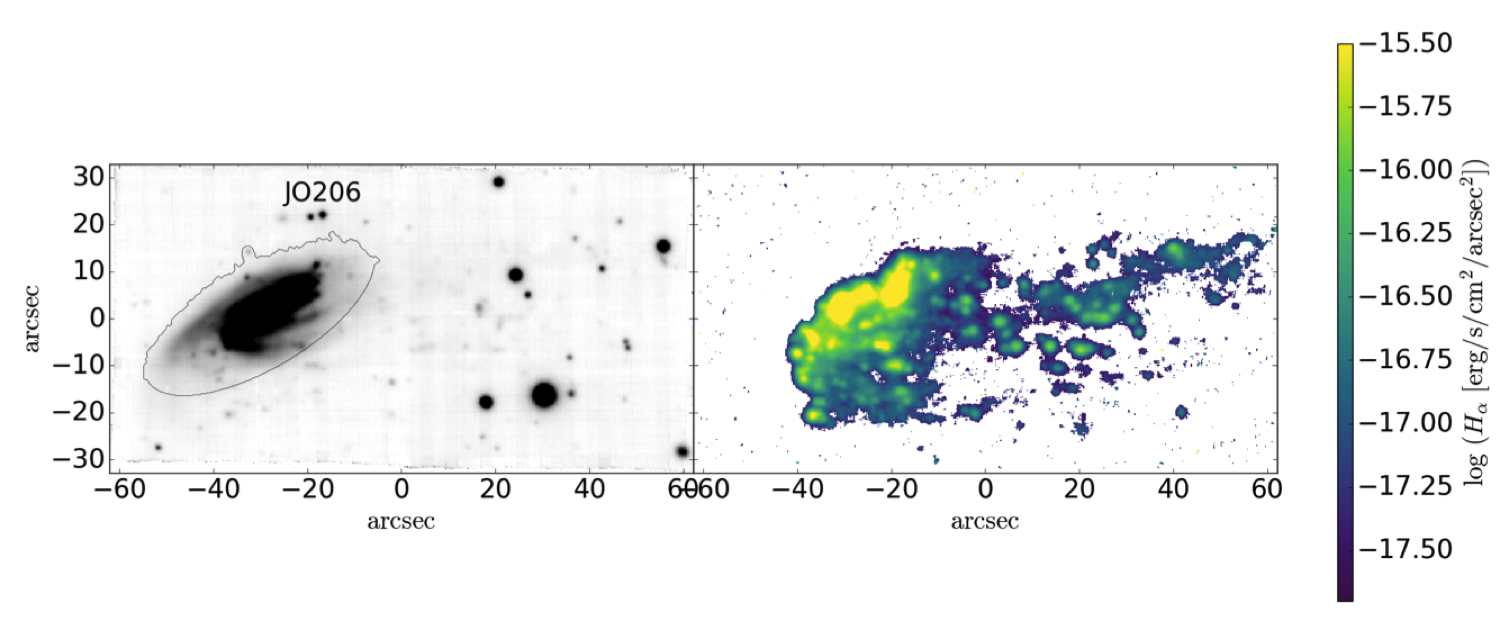}} 
\caption{Top left: Star formation rate-stellar mass relation for disks of
  jellyfish galaxies compared with undisturbed galaxies, from
  Vulcani et al. 2018b. Top right: Stellar mass distribution of star-forming
  clumps in the tails of jellyfish galaxies as solid histograms (red:
  only clumps that are star-forming according to the BPT diagrams;
  black: all clumps). For comparison, the dashed histogram is for
  clumps in the disks. Bottom: the jellyfish galaxy JO206 with its
  90kpc-long tail of $\rm H\alpha$ emitting gas (right), and its
  optical image dominated by the stellar disk (left). The star-forming
clumps stand out in the $\rm H\alpha$ image, where also the diffuse
emission is visible.}
\label{fig3}
\end{center}
\end{figure}

On a galaxy-wide scale, generally the star formation rate (SFR) in
the disks of galaxies undergoing stripping is slightly but significantly enhanced
with respect to undisturbed galaxies of similar mass, i.e. 
galaxies undergoing stripping tend to lie above the SFR-stellar mass relation (Vulcani et
al. 2018b, Fig.~4). Moreover, jellyfish galaxies follow the
mass-metallicity relation of non-stripped cluster galaxies, with
metallicities higher than field galaxies of similar mass (Franchetto et al. 2020).
Even more surprising is that new stars can form in situ in the tails of
stripped gas. This was already evident from UV studies (e.g. Smith
et al. 2010, Hester et al. 2010), and 
UV+$\rm Halpha$ studies (e.g. Boselli et al. 2018,
Abramson et al. 2011), but integral-field spectroscopy observations
have allowed us to ascertain the presence of star formation in the
tails and study its properties in an unprecedented way (Merluzzi et al. 2013, Fumagalli et
al. 2014, Fossati et al. 2016, Consolandi et al. 2017, Gullieuszik et
al. 2017, Moretti et al. 2018a, Bellhouse et al. 2019, George et al. 2018). In GASP, the
dominant ionization mechanism in the long extraplanar $\rm
H\alpha$-emitting tails is photoionization by young massive stars
(Poggianti et al. 2019a). This
star formation takes place in $\rm H\alpha$-bright, dynamically cold
star-forming clumps formed in-situ in the tails, which have $\rm
H\alpha$ luminosities typical of giant and supergiant HII regions
(e.g. like 30Dor in the LMC) and typical stellar masses 
$10^6-10^7 M_{\odot}$ (Fig.~4). Are we witnessing the formation of globular
clusters and/or Ultra Compact Dwarf galaxies? High spatial resolution
studies are needed to determine the nature and fate of these objects
(Cramer et al. 2019). The magnetic field measured for the first time
in a long jellyfish tail has been found to be highly ordered and
aligned with the tail direction. Such field, preventing heat and momentum exchange, may
be a key factor for allowing the star formation in the tails (Mueller
et al. 2020).

Another puzzle is the origin of the inter-clump, diffuse ionized emission in the
tails, which represents on average 50\% of the tail $\rm H\alpha$
emission (Poggianti et al. 2019a). The line ratios of this diffuse ionized gas (DIG) indicate
that there are areas in the tails where the ionization is powered by
SF (possibly due to photon leakage from nearby star-forming clumps,
with an average escape fraction of $\sim 18$\%), but in some cases
there is an additional (in a few cases, dominant) source of
ionization, as testified by an [OI]$\lambda$6300
excess. Most probably this is due to the interaction with the hot
intracluster medium in which the tail is embedded: either mixing, or thermal
heating or shocks give a major contribution to the tail ionization in
the jellyfish galaxy JW100 (Poggianti et al. 2019b), and this
might be the case also for other jellyfish examples for which line
ratio data is missing (Boselli et al. 2016).

\begin{figure}[b]
\begin{center}
\centerline{\includegraphics[width=2.8in]{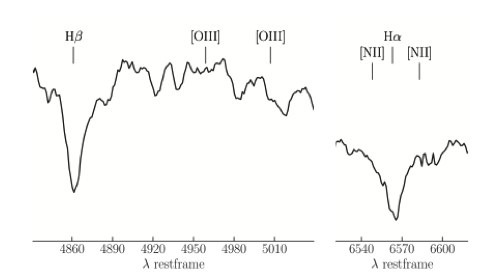}\hfill\includegraphics[width=2.8in]{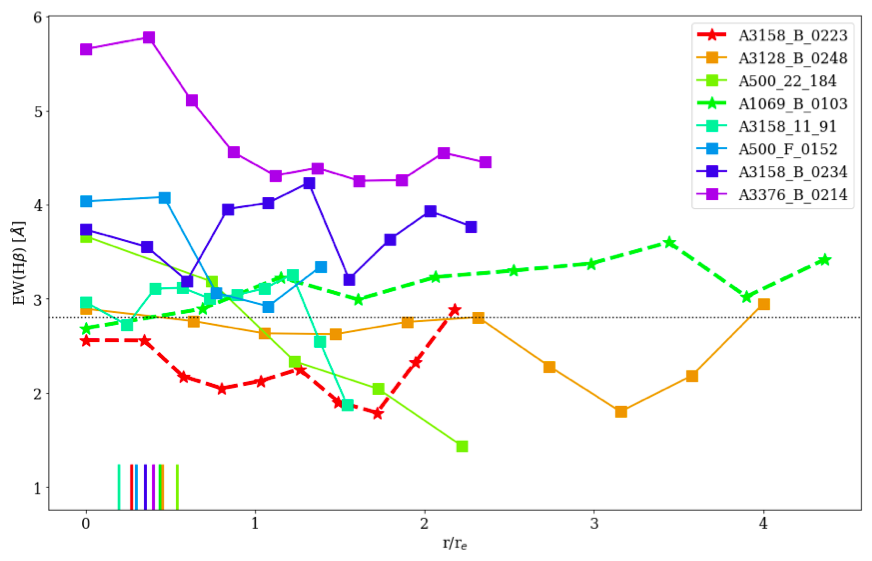}} 
\caption{Left: Strong Balmer absorption lines in the outskirts of jellyfish
  disks, where gas has already been stripped (from Gullieuszik et
  al. 2017). Right: The galactocentric radial distribution of the $\rm
  H\beta$ equivalent width in absorption in GASP
  post-starburst/post-starforming galaxies. The vertical bars at the
  bottom of the right panel indicate the size of 1 kpc in units of
  $r_e$ for each galaxy. From Vulcani et al. (2020).}
\label{fig4}
\end{center}
\end{figure}

After gas is removed by ram pressure, star formation comes to an end.
A clear signature for a recent truncation of the star formation
activity are the strong Balmer lines in absorption typical of
post-starburst/post-starforming spectra. Such a signature is present
in the outer regions of the disk of several jellyfish galaxies
(e.g. Gullieuszik et al. 2017, Poggianti et al. 2019b) and is observed throughout the disk of
those non-starforming galaxies that have recently finished to be
stripped (Vulcani et al. 2020) (Fig.~5). These are totally devoid of emission lines,
are typically located between 0.5 and 1 cluster virial radii (Owers et
al. 2019, Vulcani et al. 2020) and have been quenched outside-in (the disk outskirts first) as
expected in the ram pressure stripping scenario (Gavazzi et al. 2013).

\section{Multi-phase gas}
The number of ram pressure stripped galaxies with CO data is still rather
small, but a picture is emerging: large masses of
molecular gas have been detected both in disks and tails (Jachym et
al. 2014, 2017, 2019, Verdugo et al. 2015, Lee \& Chung 2018, 
Moretti et al. 2018b, 2020). Following the old debates about whether the molecular
gas can be stripped by ram pressure (Kenney \& Young 1989, Boselli et
al. 1997, 2014), the ALMA resolution has recently allowed to study large
individual CO clumps/complexes of $10^6-10^9
M_{\odot}$ masses of $H_2$ in the tails (Jachym et al. 2019,
Moretti et al. 2020). These studies suggest that while
the cold gas observed close to the disk may be stripped, that observed
further out in the tail forms there (see also Verdugo et al. 2015). 

\begin{figure}[b],
\begin{center}
\centerline{\includegraphics[width=3.5in]{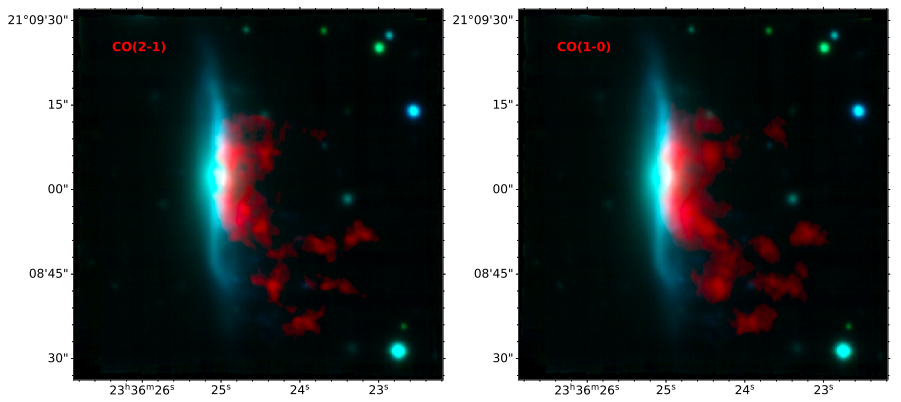}\hfill\includegraphics[width=3.2in]{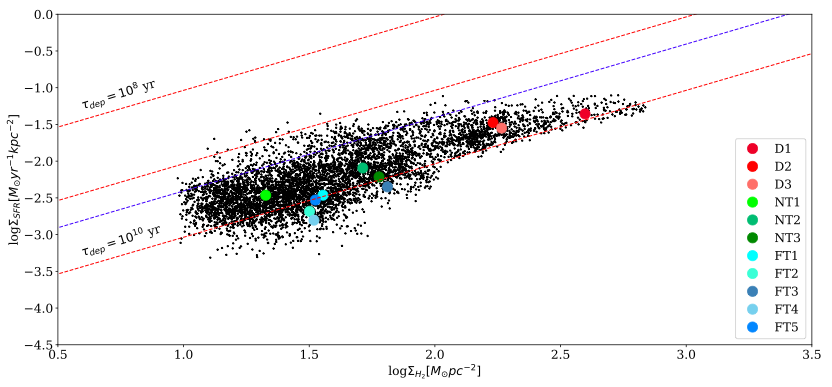}} 
\caption{Left and center: CO(2-1) and CO(1-0) emission on top of the
  I-band image of the galaxy JW100. Right: SFR surface density versus
  $H_2$ surface density (1kpc scale) for spaxels of JW100, with a few
  regions of interest in the disk and tails highlighted. In the right
  panel, the red dashed lines are fixed depletion times ($10^8$,
  $10^9$, $10^{10}$yr from top to bottom), while the blue dashed line is
  the average relation for normal nearby disk galaxies at 1 kpc
  resolution. From Moretti et al. (2020).}
\label{fig5}
\end{center}
\end{figure}

The amount of molecular gas in some jellyfishes is impressive,
($>10^9-10^{10} M_{\odot}$).
The GASP galaxy JW100 contains $2.5 \times 10^{10} M_{\odot}$ of
molecular gas (8\% of the galaxy stellar mass), of which 30\% is in
the tail (Fig.~6, Moretti et al. 2020). 
Interestingly, the CO-star
formation efficiency, defined as the ratio between the SFR and the
molecular gas mass, is low, both on the galaxy scale and on a
1kpc spatially resolved scale, yielding depletion timescales up to
$10^{10}$ yr (e.g.Vollmer et al. 2008,  Jachym et al. 2014, Verdugo et al. 2015, 
Moretti et al. 2018b, 2020, see Fig.~6).

In the tails there is a general correspondance between the spatially resolved
distribution of the various tracers related
to star formation (UV light, $\rm H\alpha$ emission and CO emission),
but it is also possible to observe directly the ``star formation
sequence'', with CO-only clumps, CO+$\rm H\alpha$+UV clumps, $\rm
H\alpha$+UV and UV-only clumps, representing the different stages of
the star formation process (Poggianti et al. 2019b).

As far as the neutral gas is concerned, HI observations paved the way to ram pressure
studies, with milestones results showing the deficiency of HI in
cluster galaxies (Haynes et al. 1984, Cayatte et al. 1990, Vollmer et
al. 2001, Chung et al. 2009, to name a few). However, the number of jellyfish galaxies with 
multiwavelength data, probing neutral, molecular and ionized gas in
the same system, is very limited, thus the origin and the
conditions allowing the presence of multi-phase tails are still
to be clarified.
Generally, when an $\rm H\alpha$ tail has been observed, sufficiently
deep HI data has also revealed a neutral gas tail. However, the
morphologies of the $\rm H\alpha$ and the HI tail can be very
different (see Fig.~7 for three example galaxies), and the kinematical
decoupling of HI and $\rm H\alpha$ can be significant (Deb et
al. 2020).  The GASP jellyfish
galaxies for which HI data is available suggest that a) during the
jellyfish phase these galaxies still possess large amounts of HI gas
(they are only slightly HI deficient, Ramatsoukou et al. 2020), but
the HI is clearly displaced from the disk, spatially and/or
kinematically, and b) there is an excess of SFR for the HI content,
compared to normal spirals, both globally and on a 1kpc scale (Fig.~8). 
In other words, the  HI-star formation efficiency (ratio of SFR over
HI mass) is higher than in normal spirals. Thus, to recap, the SFR is in excess with
respect to both the HI content and the stellar mass, but the CO
emission is in excess with respect to the SFR. As a consequence, 
in jellyfish galaxies the star formation efficiency is unusually low for
molecular gas, but unusually high for neutral gas, suggesting a very
efficient transformation of neutral into molecular gas in these
systems (Moretti et al. submitted).

We have no space here to deal with tails at yet other wavelengths
(X-ray; radio continuum), but we note that multi-$\lambda$ studies of jellyfish
galaxies including these components are growing, after the pioneering
studies of e.g. Sun et al. (2010), Gavazzi \& Jaff\'e (1985).

\begin{figure}[b]
\begin{center}
\centerline{\includegraphics[width=2.2in]{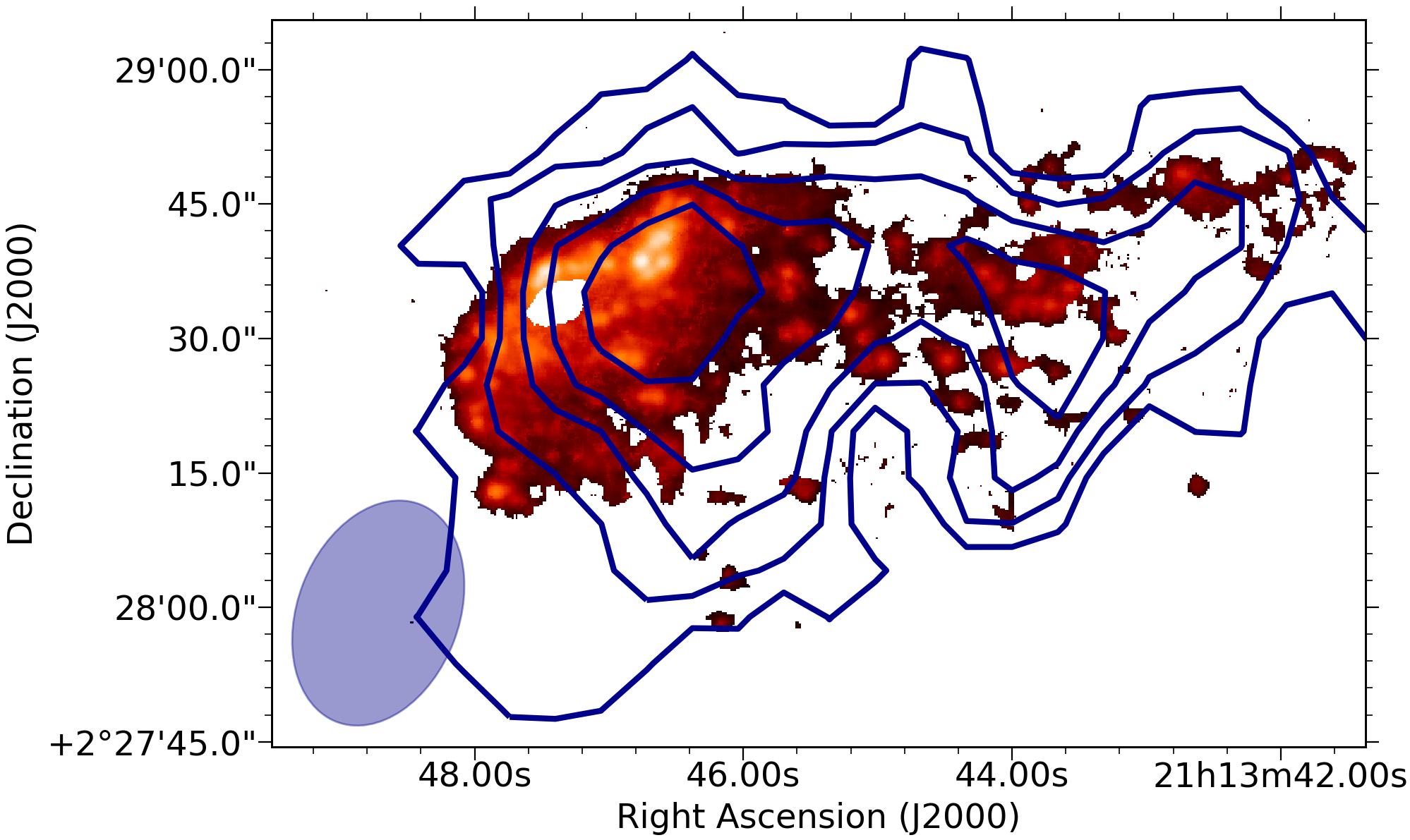}\includegraphics[width=1.4in]{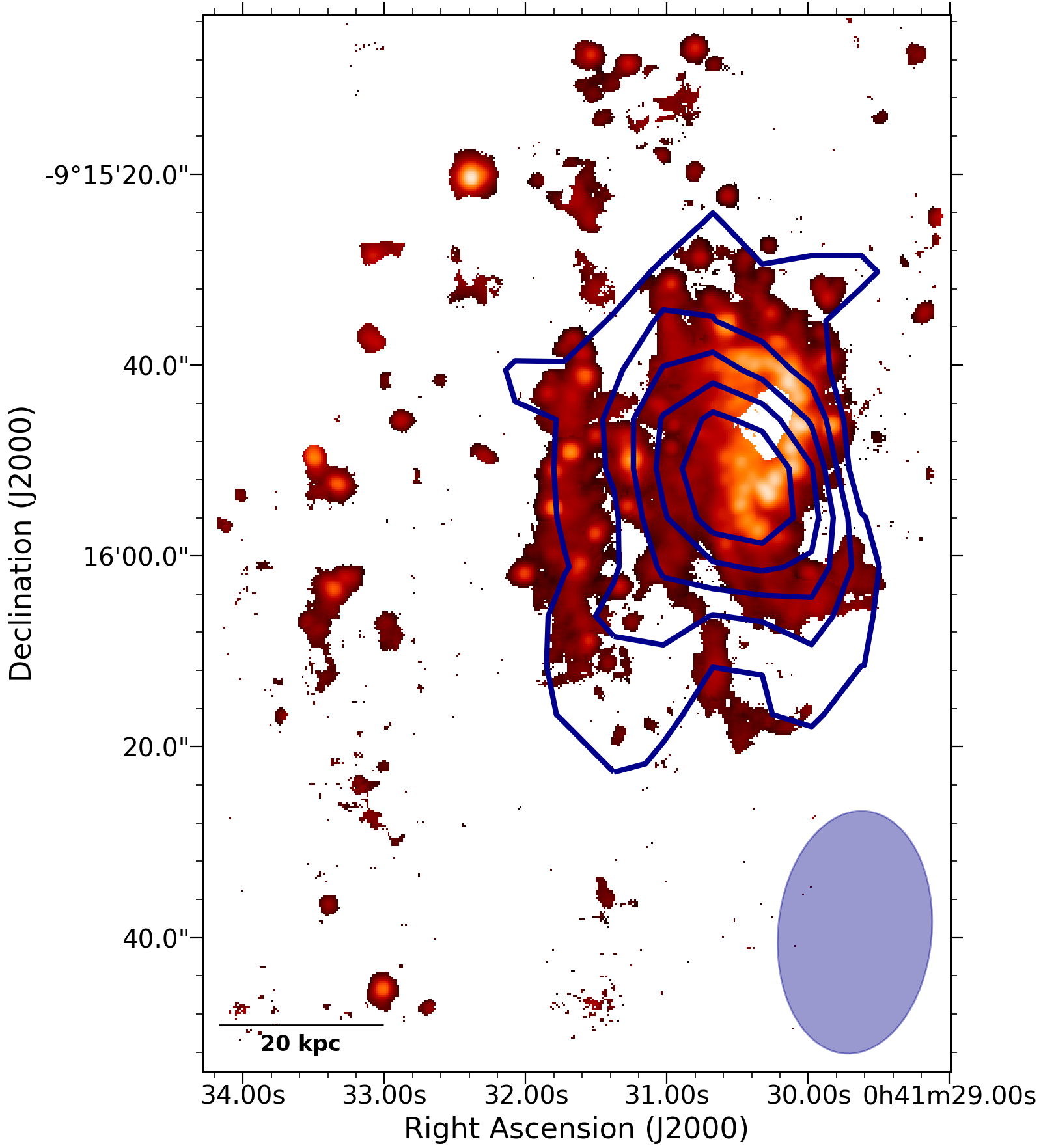}\includegraphics[width=1.6in]{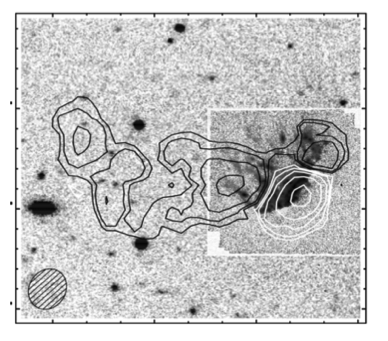}} 
\caption{HI contours on top of the MUSE $\rm H\alpha$ image (plus the
  optical broad band image outside of the inset in the right panel) of three
  GASP jellyfish galaxies (from left to right, Ramatsoku et al. 2019,
  2020 and Deb et al. 2020). In the right panel, the black contour is
  HI in emission and the white contour in absorption. The HI
  absorption is due to neutral gas along the line of sight between us
  and the central AGN. In some cases the HI and $\rm H\alpha$ tails
  are roughly co-spatial (left panel, tails are 90-kpc long), sometimes the $\rm H\alpha$ is much
  more extended than the HI (middle), and sometimes it is the opposite
(right).}
\label{fig6}
\end{center}
\end{figure}

\begin{figure}[b]
\begin{center}
\centerline{\includegraphics[width=2.5in]{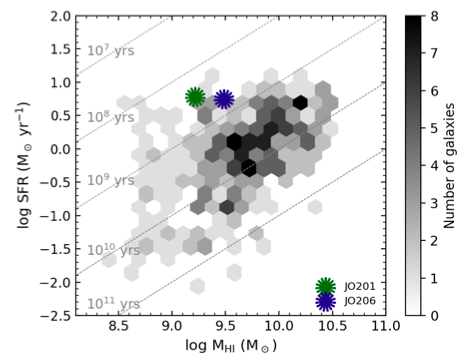}\hfill\includegraphics[width=2.5in]{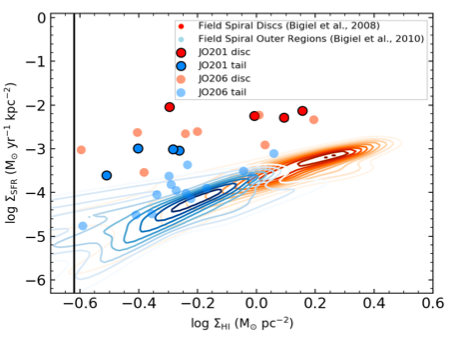}} 
\caption{Left: SFR vs HI mass for two GASP jellyfish galaxies (green
  and blue stars), compared with a control sample of spirals. Right:
  SFR and HI surface densities of the disks and tails (separately) of
  the same galaxies, compared with field spirals disks and
  outskirts. The contours in the right panel are for the inner regions
  (orange) and outer regions (blue) of field spiral galaxies, both
  convolved with the HI beam. From Ramatsoku et al. 2020.}
\label{fig7}
\end{center}
\end{figure}

To conclude, the study of ram pressure stripped galaxies is informing
us on several physical processes which are fundamental for
astrophysics in general.
We have mostly focused on the GASP results, but there is a
broad, high quality (and growing) literature on these fascinating
systems, which we hope the reader will be encouraged to explore by this
contribution of ours. We apologize for not being able to
report on the questions received after BP's talk, due to the
difficulty in hearing them remotely. BP sincerely thanks the
organizers for their kind invitation and for allowing her to give her 
presentation from the other side of the world.

\acknowledgements
This project has received funding from the European Reseach Council
(ERC)  under the Horizon 2020 research and innovation programme
(grant agreement N. 833824). Based on observations collected at the
European Organisation for Astronomical Research in the Southern 
Hemisphere under ESO programme 196.B-0578.






\end{document}